\documentclass[12pt]{article}

\usepackage[mathscr]{eucal}
\usepackage{amssymb,latexsym}
\usepackage{verbatim}
\usepackage{graphicx}
\usepackage{amsmath}
\usepackage{amsthm}
\usepackage{enumerate}
\usepackage{authblk}
\usepackage{color}
\usepackage{url}

\usepackage{caption}

\usepackage[normalem]{ulem}

\newtheorem{thm}{Theorem}[section]
\newtheorem{lem}[thm]{Lemma}
\newtheorem{Def}{Definition}[section]
\newtheorem{prop}[thm]{Proposition}

\title{A new singularity theorem for black holes which allows chronology violation in the interior  }

\author{Martin Lesourd}

\affil{University of Oxford, UK}

\begin{document}
\date{}
\maketitle
\vspace{.2in}

\begin{abstract}
The interior of the Kerr solution is singular and achronological. The classic singularity theorem by Hawking and Penrose relies on chronology, and thus does not apply to the Kerr solution. An improvement of their theorem by Kriele partially removes the requirement of chronology. However, both of these singularity theorems fail to give any information on the type or location of the incomplete geodesics. Here, using recent results of Minguzzi, we prove a new singularity theorem, specifically designed to apply to black holes, which enables locating null incomplete geodesics within the black hole interior, all the while allowing certain forms of chronology violation in the interior. 
\end{abstract}



\section{Introduction}
The current investigation lies in a line of studies aiming to better understand how singularity theorems depend on various geometric and causality conditions. A classic singularity theorem, which remained unsurpassed for a long time, is that due to Hawking and Penrose \cite{HP}; see chapter 9 of \cite{Wald} for a discussion. 
\begin{thm}[Hawking-Penrose \cite{HP}]
Let \((M,g)\) be a spacetime satisfying the timelike and null generic conditions and the strong energy condition. Suppose that \((M,g)\) either admits, a compact achronal set without edge, a trapped surface, or a point \(p\in M\) such that the expansion of the future (or past) directed null geodesics emanating from \(p\) becomes negative along each geodesic in this congruence. Then, if \((M,g)\) is chronological, there is an incomplete non-spacelike geodesic in \(M\).
\end{thm}
Theorem 1.1 relies on the assumption of chronology. There are, however, a number of explicit solutions to the Einstein field equations which contain closed timelike curves and which are still singular. Among the most physically reasonable examples is the Kerr black hole solution, the interior of which contains closed timelike curves and incomplete null geodesics; see chapter 5 of \cite{HE} for a review of the solution's properties.\\ \indent 
Thus, a question that was asked relatively early on, first by Tipler \cite{Tip} and then by Newman \cite{Newman}, was whether one could get rid of the singularities of theorem 1.1 by removing the assumption of chronology. Tipler \cite{Tip} was able to prove a singularity theorem, under a number of various different assumptions, which permitted for chronology violation. His theorem is however rather restricted in its scope of applicability and it does not provide any information on where the incomplete geodesics are. Newman's contribution \cite{Newman} was of a different kind. He constructed an explicit spacetime, based on conformal rescalings of the Godel spacetime, which violates chronology, contains a trapped surface, is non-spacelike geodesically complete, and yet satisfies all the local conditions of theorem 1.1. Newman's example shows that singularities can disappear if one totally removes chronology from theorem 1.1. However, Newman's example violates chronology in the strongest possible way. In particular, there passes a closed timelike curve through every point of his solution. \\ \indent In view of Newman's example, Kriele attempted to formulate a singularity theorem relying on a weaker form of chronology violation. Kriele \cite{Kriele} was eventually able to generalize theorem 1.1, so as to permit for certain types of chronology violation. We do not state Kriele's theorem in full here, though it is in an appendix. Our reason for doing this, as is easily verified, is that Kriele's statement of the theorem is rather long and technical. In rough, Kriele's theorem \cite{Kriele} is that of theorem 1.1 except that one replaces chronology with a strictly weaker condition permitting chronology violation. The kind of chronology violation permitted is captured in a technical statement involving various non-standard definitions. \\ \indent
One important quality that lacks in the theorem of Kriele and Hawking and Penrose concerns the location of the incomplete geodesics, and, moreover, whether they are timelike or null (or both).  \\ \indent The purpose of this paper is to prove a new singularity theorem designed specifically for black holes, which achieves things that the aforementioned ones do not. \\ \indent We formulate our theorem by positing certain geometric conditions allowing us to identify a region corresponding to a black hole interior region, though we make no mention or use of a conformal boundary in the form of `scri'. Our conclusion is that there exists incomplete null geodesics in a region corresponding to the black hole interior, even if chronology is, to some extent, violated in the interior. There are two key new ingredients which we use in this work, both of which are due to Minguzzi. \\ \indent The first is the notion of the boundary of the chronology violating region. We express the permitted chronology violation using Minguzzi's \cite{Ming3} recently offered definition of the boundary of a chronology violating set. This has the advantage of leading to a statement which relies on clear geometric conditions. In the appendix we recall Kriele's condition to show that it is, in comparison, less easy to understand. \\ \indent The second ingredient is a theorem in Lorentzian geometry, which Minguzzi proves in \cite{Ming2}. Our theorem also bears a number of interesting differences with that of Kriele, Hawking and Penrose, since in particular, we use neither the strong energy condition nor the generic conditions. Note, below, that \(p\leq q \) means \(q\in J^+(p)\), and that by \(g'\), we mean the spacetime metric \(g\) of \(M\) restricted to the open set \(O\subset M\), i.e., \(g\mid_O\).
\begin{thm}
Let \((M,g)\) be a 4-dimensional spacetime satisfying the null energy condition. Let \(V\subset M\) be an edgeless, spacelike and achronal hypersurface of co-dimension one. Let \(S\subset V\) be a compact spacelike, achronal submanifold embedded in \(V\) such that \(S\) has non-empty boundary \(\partial S\) such that \(\partial S=edge(S)=\Sigma\). Let \(N\) be a null hypersurface defined by \(N \equiv \partial I^+(S)\backslash S\) such that the following properties are obeyed:
\begin{enumerate} 
\item[(i)] \(N\cap S= \Sigma\) and for any point \(p\in N\), any point \(q\in \partial I^-(p)\cap N\) is the future endpoint of a future directed achronal null geodesic with past endpoint on \(\Sigma\);
\item[(ii)] there is no causality violation on \(N\) and the relationship ``\(\leq\)'' is closed on \(N\); 
\item[(iii)] \(\Sigma\) is a \(C^0\) surface such that for each point \(\sigma \in \Sigma\) there is only one null generator \(\eta_\sigma\) of \(N\) emanating from \(\sigma\), 
\item[(iv)] the achronal null geodesic generators of \(N\), \(\{\eta\}\) are future inextendible.
\end{enumerate}
Denote by \(O\) the open set defined by \(O=I^+(S)\) and suppose that the following properties obtain: 
\begin{enumerate}
\item[(a)] all future inextendible null geodesics in \(O\) meet a future trapped surface in \(O\),
\item[(c)] either \((O,g')\) satisfies chronology, or there is a non-empty chronology violating class \([r]\subset O\) with future and past boundary, denoted by \(B_f([r])\) and \(B_p([r])\), such that at least one of the following holds: \\ \\(i) \(B_f([r]) \cap O \neq \emptyset \), \\ \\ (ii) \((O,g')\) is non-totally vicious and either \(B_f([r])\) or \(B_p([r])\) is compact. 
\end{enumerate} 
Then there are future incomplete null geodesics in \(O\). 
\end{thm}
The conditions of the theorem are set up so as to apply to black holes which are close to being stationary, where \(N\) models the event horizon and \(V\) is a slice through the spacetime which intersects \(N\). \(S\) is a codimension one, closed, spacelike submanifold corresponding to the part of \(V\) lying inside the black hole. \(\Sigma\) is the boundary of \(S\), which thus lies on \(N\). That the event horizon is modelled by the null hypersurface \(N\) derives from certain well known properties of event horizons, in particular they are ruled by future inextendible achronal null geodesics. The specific conditions on \(N\) labelled by (i),(ii) and (iii) are supposed to capture the idea that the black hole is close to stationary. In particular, (i) and (ii) are rather standard, and (iii) models the physical idea that eventually, after things settle down, the geometry of spacetime near the horizon shoud not be so different from the stationary case. These conditions do imply that \(S\) is thought of as lying sufficiently far into the future, after things have settled down.\\ \indent Assumption (b) captures the idea that black hole interiors are filled with trapped surfaces. This assumption can be weakened in a way that preserves the contradiction of the Penrose singularity theorem. In particular, rather than requiring that all null geodesics encounter trapped surface, one could instead require that such null geodesics in \(O\) encounter either a trapped surface or a weakly trapped surface in such a way that the generic condition holds for null geodesics emanating from this surface in the direction of zero expansion. \\ \indent 
Assumption (c) permits various forms of chronology violation in the black hole interior. We have chosen to include such conditions, in view of the chronology violating and singular interior of Kerr. These conditions use the recently offered definitions and theorems of Minguzzi \cite{Ming2}, \cite{Ming3}, which are described in more detail below.

\section{Preliminaries} 
The convention adopted here follows \cite{MingSanch} so that by a spacetime \((M,g)\) we mean a \(C^{r}\), \(r\geq \{3\}\) connected, time-oriented Lorentzian manifold of dimension \(d\geq 4\). The elements in \(M\) are as differentiable as permitted by \(M\) so \(C^{r-1}\) for \(g\) and \(C^{r-3}\) for the curvature tensor \(R\). \\ \indent By null (timelike) convergence condition we mean \(R(k,k) \geq 0\) where \(k\) is any null (timelike) vector and \(R(\cdot,\cdot)\) is the Ricci tensor. \\ \indent By a \textit{future trapped surface} we mean a closed smooth compact spacelike submanifold of \(M\) of co-dimension \(2\), with null expansion scalars \(\theta_{+,-}\) associated to the future directed null normal vector fields \(l_{+,-}\) satisfying \(\theta_{+,-}<0\). A \textit{marginally trapped surface} (MTS) is the same geometrical object except that \(\theta_{+,-}\leq0\) and one of these is zero, a \textit{marginally outer trapped surface} (MOTS) has \(\theta_+=0\) and no restriction on \(\theta_-\). We say that a null geodesic \(\gamma:I\to M\) encounters a surface \(S\) if there is a point \(p\in S\) such that \(\gamma(a)=p\) for some \(a\in I\).\\ \indent  By \(x\ll y\) (\(x\leq y\)) we mean that there exists a future directed timelike (causal) curve from \(x\) to \(y\). By a future (past) null ray we mean a future (past) inextendible achronal causal curve. By a null line we mean an inextendible achronal causal curve. \\ \indent The concept of an edge is defined as follows. Let \(S\subset M\) be achronal, then \(p\in \overline{S}\) is an edge point of \(S\) provided every neighborhood \(U\) of \(p\) contains a timelike curve \(\gamma\) from \(I^-(p,U)\) to \(I^+(p,U)\) that does not meet \(S\). We denote by \(edge(S)\) the set of edge points of \(S\).\\ \indent By a totally vicious spacetime \((M,g)\) we mean one where \(I^{\pm}(p)=M\) for all \(p\in M\). \\ \indent We now recall the following standard result, cf., \cite{BBE}.
\begin{prop} 
Let \(S\) be closed. Then each \(p\in \partial I^+(S)\backslash S\) lies on a null geodesic contained in \(\partial I^+(S)\), which either has a past endpoint on \(S\), or else is past inextendible in \(M\).
\end{prop}
We shall now recall some recent results of Minguzzi, though our account merely includes what is necessary. The interested reader is encouraged to consult the comprehensive description proposed in the original papers \cite{Ming2},\cite{Ming3}.
\begin{thm}[Minguzzi \cite{Ming2}]
Let \((M,g)\) be any spacetime. 
\begin{enumerate}
\item[(a)] A chronological spacetime \((M,g)\) without future null rays is globally hyperbolic.
\item[(b)]A non-totally vicious spacetime \((M,g)\) without null rays is globally hyperbolic.
\item[(c)] A chronological spacetime \((M,g)\) without null lines is stably causal.
\item[(d)] A non-chronological spacetime \((M,g)\) without null lines is either totally vicious or has a non-empty chronology violating open set, possibly made up of distinct sets  \(C\equiv \cup_\alpha C_\alpha\). The components of the closed boundaries \(B_{\alpha k}\) of these sets \(\partial C_\alpha \equiv \cup _k B_{\alpha k}\) are all disjoint and non-compact. Finally, non-totally vicious spacetimes without null lines are non-compact.
\end{enumerate}
\end{thm}
Another ingredient we shall use, also due to Minguzzi, is the recent proposal \cite{Ming3} for the definition of the boundary of a chronology violating region.\footnote{The boundary of a chronology violating set \(C\) had been considered in the literature \cite{H}, \cite{Thorne} but these studies were based on the non-trivial assumption that \(M \backslash \overline{C}\) is globally hyperbolic (i.e. so that \(\partial C\) can be identified as a component of a Cauchy horizon). Part of the appeal of Minguzzi's analysis is that it is based only on properties that are intrinsic to chronology violation and that it makes no use of such assumptions.} The chronology violating region \(\{ C:= x\ll x\}\) is defined as the set of of points through which there passes a closed timelike curve. The relation \(x\sim y\) if \(x\ll y\) and \(y\ll x\) is an equivalence relation in \(C\), and, as is known from Carter \cite{Carter}, it splits the chronology violating region into (open) equivalence classes. We denote such a class in square brackets \([x]=I^+(x)\cap I^-(x)\). An important fact about chronology violating class is encoded in the following lemma. 
\begin{lem}[Minguzzi \cite{Ming3}]
Let \([r] \) be a chronology violating class. If \(p\in \partial [r]\) then through \(p\) passes a future or past null ray contained in \(\partial [r]\) and possibly both.
\end{lem}

Using this definition, a short argument leads to the following.
\begin{Def}
Let \([r]\) be a chronology violating class. The set \(R_f( [r] )\) is that subset of \([\partial r]\) which consists of the points \(q\) through passes a future null ray contained in \([\partial r]\). The set \(R_p([r])\) is defined analogously.
\end{Def}
\begin{lem}[Minguzzi \cite{Ming3}]
The sets \(R_p([r])\) and \(R_f([r])\) are closed and \(\partial [r] =R_p([r])\cup R_f([r])\). 
\end{lem}
After showing that a number of relations hold between \(R_f(p)([r])\), \(I^{+(-)}([r])\), \(\partial I^{+(-)}([r])\) and \(\overline{[r]}\), Minguzzi considers the sets \(B_{f(p)}([r]) \equiv \overline{[r]}/ I^{-(+)}([r])\) and proves that they have the following properties.
\begin{prop}[Minguzzi \cite{Ming3}]
The set \(B_{p(f)}([r])\) is closed, achronal, and generated by past (future) null rays. 
\end{prop}
\begin{prop}[Minguzzi \cite{Ming3}] 
The following identity holds \(\partial [r]=B_p([r])\cup B_f([r])\). 
\end{prop}
Minguzzi finally settles with the following definition for the boundary of the chronology violating class.
\begin{Def}
The sets \(B_f([r])\) and \(B_p([r])\) are respectively the future and the past
boundaries of the chronology violating class \([r]\).
\end{Def}
\section{Proof of theorem 1.1}
Let \((O,g')\) be a spacetime with \(g'=g\mid_{M\cap O}\). We shall treat \((O,g')\) as a spacetime in its own right and investigate the consequences in \((M,g)\). Firstly, we observe that if \((M,g)\) is future null complete then so is \((O,g')\).\\ \\
\textbf{Step 1}. If \((O,g')\) is chronological and future null complete, then \(N\) must be compact. \\ \\
By a standard focusing lemma, eg., proposition 9.3.9 in \cite{Wald}, (b) implies that if \((O,g')\) is future null geodesically complete, then there can be no future null rays in \(O\). A future null ray is achronal but by proposition 9.3.9, any directed null geodesic in \(O\) must meet a future trapped surface and focus. One might think that this is not enough to get rid of future null rays because we could consider the null geodesic whose endpoint is in the future of the point at which the null geodesic focused. Perhaps this new geodesic is a null ray. But this is straightforwardly seen to be impossible, for (b) applies to all null geodesics, and thus also applied to this geodesic which we have considered as a candidate ray. \\ \indent It then follows, by theorem 2.1, that either \((O,g')\) admits a Cauchy surface or else violates chronology. By the Penrose's collapse singularity theorem \cite{Pen}, given that \((O,g')\) satisfies the null energy condition, contains a closed future trapped surface and is future null geodesically complete, its Cauchy surface must be compact. \\ \indent
Recall here that \(I^+(S)=O\). By \cite{BernSanch} we can take the compact Cauchy surface for \((O,g')\) to be a spacelike slice \(V \subset O\). Consider any point \(q\in \partial O\). By definition of \(N\) and of \(S\), any future timelike curve that intersects \(N\) and \(S\) must enter \((O,g')\). A standard result of globally hyperbolic spacetimes and Cauchy surfaces then dictates that all such timelike curves, if made future inextendible, will intersect \(V\) exactly once. This permits us to define a continuous map \(\psi: \Sigma \cup \partial I^+(S) \to V \) by following integral curves of a smooth future directed timelike vector field in \(M\). Note that \(\Sigma \cup \partial I^+(S)\) and \(\Sigma\) are both closed. Continuity of \(\psi\) and compactness of \(\Sigma\) imply, together, that \(\Sigma \cup \partial I^+(S)\) is compact. The set \(\Sigma \cup \partial I^+(S)\) is also closed and by definition \(N\) is a closed subset of \(\Sigma \cup \partial I^+(S)\) and is thus also compact. \\ \\
\textbf{Step 2}. By assumptions of theorem 1.1, \(N\) is homeomorphic to \(\Sigma\times[0,\infty)\).\\ \\
Here, we shall use a variation on an argument that was exploited in \cite{BG} with somewhat different assumptions. \\ \indent We first make the following observation. By assumption, \(N\) is generated by achronal null geodesics \(\{\eta\}\) which are future inextendible and which have past endpoints on \(\Sigma\). Moreover, from our assumption on the geometry of \(N\) in the vicinity of \(S\), it follows that for each point \(\sigma \in \Sigma\), there is a unique future inextendible null geodesic generator \(\eta_\sigma\) emanating from \(\sigma\). \\ \indent
Consider the set \(A\) defined by \(A=\{a\in N: a\in \eta_\sigma, \sigma \in \Sigma \}\). \\ \indent 
We shall show that \(A\) is both open and closed in \(N\). It will then follow that \(A=N\). \\ \indent 
That \(A\) is open in \(N\). For each \(\sigma \in \Sigma\), let \(Q(\sigma)\) denote the initial null tangent vector to \(\eta_\sigma\). Our assumption that for each point \(\sigma \in \Sigma\) there is only one null generator \(\eta_\sigma\) of \(N\) emanating from \(\sigma\) guarantees that \(Q(\sigma)\) varies continuously with \(\sigma\). That is, \(\sigma\to Q(\sigma)\) is a continuous null vector field along \(\Sigma\). \\ \indent Now consider the map defined by flowing along the generators of \(N\). That is, we consider the map \(\chi:U\subset \Sigma\times [0,\infty)\to N\) defined by \(\chi(s,\lambda)=exp_\sigma(\lambda Q(\sigma))=\eta_\sigma(s)\). The set \(U\) represents the largest set in \(\Sigma\times [0,\infty)\) on which \(\chi\) is defined, which may be strictly smaller. The continuity of \(\sigma \to Q(\sigma)\) and the behavior of geodesics guarantees that \(\chi\) is continuous. Since the generators of \(N\) cannot intersect, \(\chi\) is one to one. Hence it follows from invariance of domain \cite{Munk} that \(A=\chi(U)\) is open.  \\ \indent 
That \(A\) is closed in \(N\). This follows by the assumption noted in the previous paragraph and the compactness of \(S\). In particular, let \(\{p_n\}\) be a sequence of points in \(A\) such that \(p_n\to p\in N\). For each \(n\), \(p_n\) is on \(\eta_{q_n}\) for some \(q_n\in \Sigma\). Since \(\Sigma\) is compact, there exists a convergent subsequence \(q_n\to q\in \Sigma\). Since \(p_n\geq q_m \), and the causal relationship ``\(\geq\)'' is closed by assumption, it follows that \(p\geq q\). Since \(A\) is achronal, we must then have \(p\in \eta_q\) and hence \(q\in A\). Hence \(A\) is closed in \(N\).\\ \indent 
We now have \(A=N\) and \(\chi:U\subset \Sigma \times [0,\infty)\to N\) is a homeomorphism onto \(N\). We fix a complete Riemannian metric \(h\) in \(M\). For each \(s\in \Sigma\), reparametrize \(\eta_\sigma\) with respect to arclength in the metric \(h\). Since \(\eta_\sigma\) is future inextendible, the reparametrized curve \(\tilde{\eta_s}\) will be defined on \([0,\infty)\). It follows that the map \(\chi:\tilde{\Sigma}\times [0,\infty)\to H\) defined by \(\tilde{\chi}(\sigma,\lambda)=\tilde{\eta_\sigma}(\lambda)\) is a homeomorphism since \(\chi:U\to N\) is. Hence, \(N\) is homeomorphic to \(\Sigma\times [0,\infty)\), with the homeomorphism given by the null generators (suitably parametrized). \\ \\ 
\textbf{Step 3}. There is chronology violation in \((O,g')\) as permitted in theorem 1.1.\\ \\ The statements obtained in steps 1 and 2 can only be consistent if either \((O,g')\) is future null incomplete or there is chronology violation in \((O,g')\). Assuming that \((O,g')\) is future null complete, it follows that there is a non-empty chronology violating region \(Q \subseteq O\), to which can be associated a chronology violating class, which we denote by \([r]\). \\ \indent Using definition 2.1 and proposition 2.5, future null completeness along with assumption (b) imply that \(O\) contains no null rays, and, therefore, that \(B_f([r]) \cap O =\emptyset \). This leads to condition (c)-i. \\ \indent For (c)-ii, recall that lack of future null rays implies the lack of null lines, and thus, by theorem 2.2, either \((O,g')\) is totally vicious, or the boundaries of the chronology violating class is non-compact. Assuming non-total viciousness of \((O,g')\), compactness of either one of the boundaries provides the relevant contradiction. This completes the proof of theorem 1.1.

\section{Discussion} 
The strongest assumption of theorem 1.1 is (c), i.e., that all future directed null geodesics future inextendible in \(O\) meet future trapped surfaces in \(O\). This captures the idea that a black hole interior is a region of strong gravity and that the trapped surfaces are inside the event horizon. There are a number of classic theorems justifying the latter in various contexts of asymptotic predictability, eg., \cite{HE}. As for the former, it is difficult to find a way of providing a mathematically rigorous justification, other than basing it on existing black hole solutions. 
\\ \indent 
On this point, there is also the classic result of Schoen and Yau \cite{SY} showing that initial data sets with sufficiently dense concentrations of matter contain MOTS. Further such existence results for trapped surfaces and or weakly trapped surfaces would be welcome. On this issue, we note the results in \cite{ChruscielGall} showing that any initial data containing a MOTS which lies in an region satisfying a no-KID condition, is arbitrarily close to another initial data in which there is a strictly outer trapped surface.\\ \indent
It would be preferable to possess a number of illustrative examples for theorem 1.1. We have in mind the Kerr-Newman class, since some of these solutions describe black hole interiors which contain both closed timelike curves and singularities. The issue with the Kerr-Newman class is that those solutions are usually considered in their maximal analytic form. In this form, it is possible to access another universe by going into the black hole interior. This feature makes the theorem inapplicable to these full extensions, because \(O=I^+(S)\) extends into different universes, which are regions going beyond the regions of applicability of the theorem. We plan to address this particular issue in further work.  \\ \\ 
\textbf{Acknowledgements} I would like to thank the Ruth and Nevill Mott Scholarship and the AHRC for funding this research. I would also like to thank Eric Ling, Ettore Minguzzi, Aron Wall and an anonymous referee for comments regarding preliminary versions of this article.

\newpage
\section*{Appendix - Kriele's theorem}
To define the chronology violation permitted in Kriele's theorem, we first need the following definition. 
\begin{Def}
Let \((M,g)\) be a spacetime and \(\mathcal{S}\) be a compact set that is achronal in some neighborhood \(U\). We denote by \(\beta_p\) the generator of \(E^+(\mathcal{S},U)\) through \(p\) prolonged to the future as far as possible without encountering a focal point. Then \(e^+(\mathcal{S},M)\equiv \overline{ q\in M \backslash q\in \mathcal{S}, q= \beta_p(t)}\) is called the generalized future horismos of \(\mathcal{S}\). \(cl(\mathcal{S},M)\equiv q\in M\) such that \(q\) is not the future endpoint of some generator of \(e^+(\mathcal{S},M)\), is a set called the cut locus of the generalized future horismos of \(\mathcal{S}\).
\end{Def}
\begin{thm}[Kriele \cite{Kriele}]
Theorem 1.1 remains valid if chronology is replaced with the following condition. Neither \(cl(\mathcal{S},M,+)\) nor any \(cl(\mathcal{D},M,-)\), where \(\mathcal{D}\) is a compact topological submanifold (possibly with boundary) with \(\mathcal{D}\cap \mathcal{S}\neq 0\) contains a causal curve which can be rendered nontrivial and closed by an arbitrary small deformation of a piece of it.
\end{thm}
Kriele's theorem has the virtue of being applicable to a wide range of situations. Its condition on chronology violation is, however, somewhat harder to understand than that used in theorem 1.2. 

\end{document}